\documentclass[english]{article}
\usepackage[T1]{fontenc}
\usepackage[utf8]{inputenc}
\usepackage{lmodern}
\usepackage{color}
\usepackage[svgnames]{xcolor}
\usepackage{bm}
\usepackage{amsmath}
\usepackage{amssymb}
\usepackage{authblk}

\makeatletter

\usepackage{color}
\usepackage{bm}
\usepackage{amsmath}
\usepackage{amssymb}
\usepackage{booktabs}
\usepackage{siunitx}
\usepackage{float}
\usepackage{array}
\usepackage{placeins}
\makeatletter
\usepackage{epsfig}\usepackage{subfigure}\usepackage{amsfonts}\usepackage{array}
\usepackage{theorem}

\newcommand{\bml}{\begin{subequations}}
\newcommand{\eml}{\end{subequations}}
\allowdisplaybreaks[1]

\newcommand{\qed}{\nobreak \ifvmode \relax \else
      \ifdim\lastskip<1.5em \hskip-\lastskip
      \hskip1.5em plus0em minus0.5em \fi \nobreak
      \vrule height0.75em width0.5em depth0.25em\fi}
\usepackage{vmargin}
\setpapersize{custom}{8.5in}{11.0in}
\setmarginsrb{1.0in}{1.0in}{1.0in}{1.0in}%
            {0pt}{10mm}{0pt}{0mm}

\let\originalleft\left
\let\originalright\right
\renewcommand{\left}{\mathopen{}\mathclose\bgroup\originalleft}
\renewcommand{\right}{\aftergroup\egroup\originalright}

\usepackage{mathtools}

\usepackage{multirow}
\makeatother

\usepackage{babel}
\usepackage{caption}
\usepackage{tikz}

\usepackage{pifont}

\usepackage{dcolumn}
\newcolumntype{d}[1]{D{.}{.}{#1}}
\usepackage{siunitx}
\usepackage{url}
\usepackage{caption}
\usepackage[utf8]{inputenc}
\usepackage[T1]{fontenc}
\usepackage{hyperref}
\usepackage{csquotes}
\usepackage[
  backend=biber,
  style=ieee,
  sorting=none,
  maxnames=3,
  url=true,
  doi=true,
]{biblatex}

\usepackage{listings}
\usepackage{xcolor}   

\lstdefinelanguage{text}{
	morekeywords={},      
	sensitive=false,      
	morecomment=[l]{//},  
	morestring=[b]"      
}

\lstdefinestyle{mystyle}{
	basicstyle=\ttfamily\small,    
	numbers=left,
	numberstyle=\tiny\color{gray},
	stepnumber=1,
	numbersep=5pt,
	frame=single,
	breaklines=true,
	breakatwhitespace=true,
	tabsize=2,
	showspaces=false,
	showstringspaces=false,
	showtabs=false,
	keywordstyle=\color{blue},
	commentstyle=\color{gray},
	stringstyle=\color{red!60!black},
	captionpos=t
}

\usepackage{listings}
\usepackage{float}                   
\newfloat{listing}{htbp}{lop}[section]
\floatname{listing}{Listing}

\begin{document}
\title{Higher-Order Automatic Differentiation Using Symbolic Differential Algebra : Bridging the Gap between Algorithmic and Symbolic Differentiation}
\author[1]{He Zhang}
\affil[1]{Thomas Jefferson National Accelerator Facility, Newport News, VA, USA}
\date{}
\maketitle

\begin{abstract}
  In scientific computation, it is often necessary to calculate higher-order derivatives of a function. Currently, two primary methods for higher-order automatic differentiation exist: symbolic differentiation and algorithmic automatic differentiation (AD). Differential Algebra (DA) is a mathematical technique widely used in beam dynamics analysis and simulations of particle accelerators, and it also functions as an algorithmic automatic differentiation method. DA automatically computes the Taylor expansion of a function at a specific point up to a predetermined order and the derivatives can be easily extracted from the coefficients of the expansion. We have developed a Symbolic Differential Algebra (SDA) package that integrates algorithmic differentiation with symbolic computation to produce explicit expressions for higher-order derivatives using the computational techniques of algorithmic differentiation. Our code has been validated against existing DA and AD libraries. Moreover, we demonstrate that SDA not only facilitates the simplification of explicit expressions but also significantly accelerates the calculation of higher-order derivatives, compared to directly using AD.
\end{abstract}

\section{Introduction}
Calculating higher-order derivatives of a given function often plays an important role in numerical simulations across many fields, \textit{e.g.}, physical, chemical, or biological system simulations \cite{majji2008high, shu2003high, abbott2021arbitrary}, optimization and control \cite{enciu2010automatic}, and machine learning and artificial intelligence \cite{rajeswaran2019metalearningimplicitgradients, baydin2018automatic}. Currently, there are two main methods for higher-order automatic differentiation: symbolic differentiation and algorithmic differentiation.
Symbolic differentiation works by taking a mathematical expression and applying the rules of calculus, such as the power rule, product rule, and chain rule, to obtain its derivatives. The advantage of symbolic differentiation is that it provides an explicit expression for the derivatives. However, the computational complexity and time may increase significantly when calculating higher-order derivatives. It may also suffer from the problem of "expression swell", \cite{corliss1988applications} generating unnecessarily large expressions, which are unsuitable for numerical computation.
Algorithmic differentiation (AD), on the other hand, does not rely on the specific form of the function. As long as the function's value can be computed, its derivatives can be calculated to machine precision. This makes AD widely used in numerical computations. Automatic differentiation is probably the most popular algorithmic differentiation method, the common types of which include forward mode differentiation, reverse mode differentiation, and mixed mode differentiation. \cite{nolan1953analytical, wengert1964simple,speelpenning1980compiling,rall2006perspectives}

The differential algebra (DA), also referred to as truncated power series algebra (TPSA), is a method widely used in accelerator physics, typically to generate the high-order transfer map of one or more particle accelerator beamline components for beam dynamics study.\cite{abqtpsa, pada, cpo3ho, cpo4berz} It can also be used in more sophisticated methods, \textit{e.g.}, symplectic tracking \cite{caprimap}, normal form analysis \cite{monthnf}, verified integration \cite{rdaint}, global optimization \cite{makino2005verified}, and fast multipole method for pairwise interactions between particles \cite{FMMCPO2010}. DA automatically calculates the Taylor expansion of a function at a specific point. The higher-order derivatives of the function can be extracted from the coefficients of the respective terms in the Taylor expansion. Hence DA provides an alternative algorithmic way for higher-order differentiation. 

In this paper, we present an innovative method that combines algorithmic differentiation with symbolic calculation, \textit{i.e.}, the symbolic differential algebra (SDA), for higher-order differentiation.  SDA combines the strengths of both algorithmic differentiation and symbolic differentiation, producing explicit expressions for higher-order derivatives through the computational methods of algorithmic differentiation. We validate SDA with existing DA and AD libraries. We also demonstrate that it is easy to simplify the explicit expressions using SDA, and it is much faster to calculate the higher-order derivatives using the explicit expressions obtained by SDA than to calculate them using AD directly.

\section{Differential algebra}\label{sec:da}
In this section, we give a very brief introduction on DA/TPSA from a practical computational perspective. A more complete treatment is found in \cite{AIEP108book} and \cite{chao2022special}.

The fundamental concept in DA is the DA vector. To make this concept easier to understand, we can consider a DA vector as the Taylor expansion of a function at a specific point.
Considering a function $f(\mathbf{x})$ and its Taylor expansion $f_{\mathrm{T}}(\mathbf{x}_0)$ at the point $\mathbf{x}_0$ up to the order $n$, we can define an equivalence relation between the Taylor expansion and the DA vector as follows

\begin{equation}
  [f]_n = f_{\mathrm{T}}(\mathbf{x}_0) = \sum {C_{n_1,n_2, ..., n_v}} \cdot d_1^{n_1} \cdot \dots \cdot d_v^{n_v}, \label{eq:da}
\end{equation}
where $\mathbf{x} = (x_1, x_2, \dots, x_v)$, and $n \ge n_1 + n_2 + \dots + n_v$. Here $d_i$ is a special number: it represents a small variance in $x_i$. Generally one can define a DA vector by directly setting values to respective terms, without defining the function $f$. The addition and multiplication of two DA vectors can be defined straightforwardly. To add two DA vectors, we simply add the coefficients of the like terms. To multiply two DA vectors, we multiply each term in the first one with all the terms in the second one and combine like terms while ignoring all terms above order $n$. So given two DA vectors $[a]_n$ and $[b]_n$ and a scalar c, we have the following formulae:

\begin{eqnarray}
  [a]_{n}+[b]_{n} & := & [a+b]_{n},\nonumber \\
  c\cdot[a]_{n} & := & [c\cdot a]_{n},\label{eq:addmlt}\\
  {}[a]_{n}\cdot[b]_{n} & := & [a\cdot b]_{n},\nonumber 
\end{eqnarray}

According to the fixed point theorem \cite{AIEP108book}, the inverse of a DA vector that is not infinitely small can be calculated iteratively in a limited number of iterations. Once the fundamental operators are defined, the DA vector can be used in calculations just as a number. Combining addition, multiplication, subtraction, division and power series, almost all functions that can be represented on a computer can be carried out on DA vectors. Fundamental functions, such as the exponential, the logarithm, the trigonometric, and the hyperbolic functions are usually supported by DA calculation packages. 

It is easy to see from Eq.~(\ref{eq:da}) that a derivative of the function $f(\mathbf{x})$ with respect to its variables $x_1, x_2, \cdots, x_v$ at a given point $\mathbf{x}_0$ can be extracted from the respective coefficient in the DA vector:
\begin{equation}
  \frac{\partial^{n_\mathrm{o}}f}{\partial^{n_1}x_1,\partial^{n_2}x_2, \cdots, \partial^{n_v}x_v} = n_1!n_2!\cdots n_v!\cdot C_{n_1,n_2, ..., n_v},
\end{equation}
where $n_\mathrm{o} = n_1+n_2+\cdots +n_v$ and  $n_i!$ is the factorial of $n_i$. The numerical DA actually provides a way to calculate the derivatives for any given function as long as the function can be carried out using the fundamental operators in computer no matter having an explicit expression or not.

\section{Symbolic differential algebra}\label{sec:sda}
The SDA combines the DA technique and the symbolic calculation technique. The SDA vector and the operators are defined in  exactly the same way as in DA. The only difference between them is that the $\mathbf{x}_0$ in $f_{\mathrm{T}}(\mathbf{x}_0)$ has to be a number or numbers in DA while it is a symbol or symbols in SDA. The coefficients in an SDA vector are expressions of the variables, defined as symbols, in the function $f$. Instead of the derivative value at   a specific $\mathbf{x}_0$, one can obtain an explicit expression for the derivative in term of $\mathbf{x}_0$ from the respective coefficient in an SDA vector. The expression can be used to calculate the derivatives for all valid values of $\mathbf{x}_0$.  

We have developed a C++ package for SDA calculation \cite{zhang2024SDA} based on the DA package, cppTPSA \cite{zhang2024cpptpsa}, and the symbolic calculation package, SymEngine \cite{Fernando2024SymEngine}. To the best of our knowledge, this is the first and currently the only symbolic DA package in the world. In the following, all the SDA calculations are carried out using this package and all the DA calculations are carried out using the cppTPSA package. Listing~\ref{code:sda} shows C++ style pseudocode as an example to demonstrate how the DA and SDA libraries work. In this example, we calculate the Taylor expansion of the inverse of $r$ where $r = \sqrt{x^2 + y^2 + z^2}$. This function is often seen in scientific simulations. Line 8 calculates the Taylor expansion at $\mathbf{x}_0 = (1,1,1)$ as a DA vector, where $\mathrm{da}[i]$ with $i=0,1,2$ is the base in the respective direction.  In lines 10 - 11, we define three symbols, x, y, and z, and calculate the symbolic Taylor expansion using them. We calculate the Taylor expansions up to the 3rd order using both the numerical DA and the symbolic DA. The numerical result is shown in Listing~\ref{output:da_sample}. The integers in the first column under "I" are simply the sequence numbers of each value. The numbers in the second column under "V [15]" are the coefficients of each term in the Taylor expansion. 15 identifies the memory slot of the current DA vector in the whole DA vector pool, which is not important in the discussion here.  Each group of integers in the third column under "Base" tells the order of the respective variable in each monomial in the Taylor expansion. In this example, we have three variables x, y, and z, hence there are three integers in each group. The value $i,j,k$ refers to the monomial $x^iy^jz^k$ in the Taylor expansion, and the value in the second column is the coefficient of this monomial. The last column under [20 / 20] shows the index of each term in the DA vector, which starts from zero in the C programming language style. The first 20 in the square brackets means there are 20 non-zero terms in this DA vector. The second 20 means there are at most 20 terms in this DA vector with three variables up to the third order. They are both 20 because this DA vector is a full vector without any zero terms. For a sparse DA vector, the first number is smaller than the second and the integers in this column are not consecutive because the zero-value terms are not shown. The symbolic result is shown in Listing~\ref{output:sda_sample}. The first column shows the orders of the bases, the second column shows the index of each term in the SDA vector, and the last column shows the coefficients of each term. Each coefficient of the SDA vector is an explicit expression in $x$, $y$, and $z$ and is valid for any reasonable choice of $x$, $y$, and $z$. Of course, the usage of both the DA and the SDA is not limited to this specific function of $1/r$. In principle, we can change the definition of the function, in lines 1 to 4 in Listing~\ref{code:sda}, as long as it can be carried out by the fundamental operators that are overloaded for the DA and SDA datatypes. This is possible even if we do not have an explicit expression for the function.  

\begin{listing}
	\centering
\begin{minipage}{0.75\textwidth}
\begin{lstlisting}[style=mystyle, language=C++, caption={Sample code for DA and SDA}, label={code:sda}]	
template <typename T>
T fun(T x, T y, T z) { return 1/sqrt(x*x + y*y + z*z); }

...

DA r_inv = fun(1+da[0], 1+da[1], 1+da[2]);

SymEngine::Expression sx("x"), sy("y"), sz("z");
SDA sr_inv = fun(sx+sda[0], sy+sda[1], sz+sda[2]);
\end{lstlisting}
\end{minipage}
\end{listing}

\begin{listing}
	\centering
\begin{minipage}{0.65\textwidth}
\begin{lstlisting}[style=mystyle, language=text, caption={A numerical DA vector}, label={output:da_sample}]
   I          V [15]              Base  [ 20 / 20 ]
  -------------------------------------------------
    1   5.773502691896258e-01     0 0 0     0
    2  -1.924500897298753e-01     1 0 0     1
    3  -1.924500897298753e-01     0 1 0     2
    4  -1.924500897298753e-01     0 0 1     3
    5   8.012344526598184e-18     2 0 0     4
    6   1.924500897298753e-01     1 1 0     5
    7   1.924500897298753e-01     1 0 1     6
    8   8.012344526598184e-18     0 2 0     7
    9   1.924500897298753e-01     0 1 1     8
   10   8.012344526598184e-18     0 0 2     9
   11   4.276668660663895e-02     3 0 0    10
   12  -6.415002990995845e-02     2 1 0    11
   13  -6.415002990995845e-02     2 0 1    12
   14  -6.415002990995845e-02     1 2 0    13
   15  -3.207501495497922e-01     1 1 1    14
   16  -6.415002990995845e-02     1 0 2    15
   17   4.276668660663895e-02     0 3 0    16
   18  -6.415002990995845e-02     0 2 1    17
   19  -6.415002990995845e-02     0 1 2    18
   20   4.276668660663895e-02     0 0 3    19
\end{lstlisting}
\end{minipage}
\bigskip
\begin{lstlisting}[style=mystyle, language=text, caption={A symbolic DA vector}, label={output:sda_sample}]
 Base  [ 20 / 20 ]        V [15]               
-------------------------------------------------
 0 0 0     0    1.0/sqrt(x**2 + y**2 + z**2)
 1 0 0     1    -1.0*x/(x**2 + y**2 + z**2)**(3/2)
 0 1 0     2    -1.0*y/(x**2 + y**2 + z**2)**(3/2)
 0 0 1     3    -1.0*z/(x**2 + y**2 + z**2)**(3/2)
 2 0 0     4    1.5*x**2/(sqrt(x**2 + y**2 + z**2)*(2*x**2*y**2 + 2*x**2*z**2 + 2*y**2*z**2 + x**4 + y**4 + z**4)) - 0.5*(x**2 + y**2 + z**2)**(-3/2)
 1 1 0     5    3.0*x*y/(sqrt(x**2 + y**2 + z**2)*(2*x**2*y**2 + 2*x**2*z**2 + 2*y**2*z**2 + x**4 + y**4 + z**4))
 1 0 1     6    3.0*x*z/(sqrt(x**2 + y**2 + z**2)*(2*x**2*y**2 + 2*x**2*z**2 + 2*y**2*z**2 + x**4 + y**4 + z**4))
 0 2 0     7    1.5*y**2/(sqrt(x**2 + y**2 + z**2)*(2*x**2*y**2 + 2*x**2*z**2 + 2*y**2*z**2 + x**4 + y**4 + z**4)) - 0.5*(x**2 + y**2 + z**2)**(-3/2)
 0 1 1     8    3.0*y*z/(sqrt(x**2 + y**2 + z**2)*(2*x**2*y**2 + 2*x**2*z**2 + 2*y**2*z**2 + x**4 + y**4 + z**4))
 0 0 2     9    1.5*z**2/(sqrt(x**2 + y**2 + z**2)*(2*x**2*y**2 + 2*x**2*z**2 + 2*y**2*z**2 + x**4 + y**4 + z**4)) - 0.5*(x**2 + y**2 + z**2)**(-3/2)
 3 0 0    10    1.5*x/(sqrt(x**2 + y**2 + z**2)*(2*x**2*y**2 + 2*x**2*z**2 + 2*y**2*z**2 + x**4 + y**4 + z**4)) - 2.5*x**3/((x**2 + y**2 + z**2)**(3/2)*(2*x**2*y**2 + 2*x**2*z**2 + 2*y**2*z**2 + x**4 + y**4 + z**4))
 ... 
 0 0 3    19    1.5*z/(sqrt(x**2 + y**2 + z**2)*(2*x**2*y**2 + 2*x**2*z**2 + 2*y**2*z**2 + x**4 + y**4 + z**4)) - 2.5*z**3/((x**2 + y**2 + z**2)**(3/2)*(2*x**2*y**2 + 2*x**2*z**2 + 2*y**2*z**2 + x**4 + y**4 + z**4))
\end{lstlisting}
\end{listing}

\FloatBarrier
\section{Higher-order automatic differentiation}

The proposed method for higher-order automatic differentiation is composed of two steps. First, given a function or a piece of code, use SDA to obtain the explicit polynomial expressions for all the derivatives. Second, parse the SDA results to generate code for the derivative calculation in the target programming language and compile the code into executables.  With the help of generative AI tools like ChatGPT \cite{OpenAIChatGPT, roumeliotis2023chatgpt}, this task can be completed without much programming experience. Listing~\ref{code:ad} shows a simple C++ function to calculate one derivative of $1/r$ based on the SDA output shown in Listing~\ref{output:sda_sample}. It takes the values of \textit{x}, \textit{y}, \textit{z}, and a string s as the arguments. The string s, which shows the orders of the bases, is mapped to an integer, which is then used as the control expression in a switch statement to select the expression for the target derivative. 

\begin{listing}
	\centering
  	\begin{minipage}{0.85\textwidth}
\begin{lstlisting}[style=mystyle, language=C++, caption={Sample code generated from SDA result}, label={code:ad}]
  static map<string, int> orders = {
	  {"000", 0},
	  {"100", 1},
	  {"010", 2},
      ... 
  }
  double derivative(std::string& s, double x, double y, double z){
    switch(orders[s]) {
        case 0: return 1/sqrt(x*x+y*y+z*z);
        case 1: return -x/pow(x*x+y*y+z*z, 3/2);
        case 2: return -y/pow(x*x+y*y+z*z, 3/2);
        ...
    }
  }
\end{lstlisting}
\end{minipage}
\end{listing}
\vspace{1em}

Both DA  and SDA provide methods for calculating derivatives of any order. The advantage of the proposed method compared to numerical DA is better efficiency. DA calculations are based on DA vectors, which can be slow, especially when the inverse of a DA vector is involved, and hence iterations have to be implemented. DA calculations cannot meet the efficiency requirements of high-performance computing in some scenarios. Using the proposed method, we avoid the time-consuming DA vector calculations  by reducing them to calculations on numbers and eliminating the iterations, thereby improving efficiency.

Another advantage of using SDA is that we can easily further simplify the expression by defining a new symbol. Again, take the function $1/r$ as an example. If we calculate in two steps, the first step would be to calculate $r^2$,
and the result would be as shown in Listing~\ref{output:sda_r2}. As expected, $r^2$ only contains terms up to the second order, although the cut-off order in the SDA calculation is set to be three. If we define a new symbol $r$, we can replace the coefficient of the constant part (the first term with base order 0 0 0) in the SDA vector using $r^2$. After we carry out the second step $1/\sqrt(r^2)$, the final result of $1/r$ is shown in Listing~\ref{output:sda_r_inv}.  Compared with Listing~\ref{output:sda_sample}, the expressions are significantly simplified, especially for the higher-order terms. Moreover, due to the reduced number of terms in the SDA vector during intermediate steps, the calculation process is also simplified accordingly. 
In Table~\ref{tab:sda_terms}, we list the number of terms when expressing the derivatives as a function of three variables, $f(x,y,z)$, or of four variables, $f(x,y,z,r)$. To explain how we calculate the number of terms of each derivative, consider the second derivative with respect to $z$, the line starting with "0 0 2" in Listing~\ref{output:sda_sample}, as an example. The corresponding expression can be initially decomposed into two major components, separated by a minus sign. In the second component, a summation over three terms is required. For the first component, the expression involves the multiplication of two parenthetical expressions. The first parenthesis contains a summation of three terms, while the second parenthesis contains a summation of six terms. Consequently, the total number of terms in this expression is $2+3+3+6=14$. In contrast, the corresponding expression in Listing~\ref{output:sda_r_inv} contains only two terms. It should be noted that this analysis does not account for the number of operations, typically multiplication or division, within each term; thus, the complexity of the expression is only approximately represented. The first column in Table~\ref{tab:sda_terms} shows the order of the derivative. The second column $N_\mathrm{d}$ shows the number of all the partial derivatives with this order. The column $N_\mathrm{t}$ shows the total number of terms for all the derivatives with the respective order. The column Max, Min, and Avg show the maximum, minimum, and average number of terms in one derivative with this order.   From the data in Table~\ref{tab:sda_terms}, we can see that when three variables are used, the number of terms is generally proportional to the square of the order. However, with the introduction of a fourth variable, the number of terms increases approximately linearly with the order. Therefore, using four variables when dealing with higher orders can significantly reduce the number of terms and enhance computational speed.

\vspace{1em}
\begin{listing}[H]
	\centering
\begin{minipage}{0.55\textwidth}
\begin{lstlisting}[style=mystyle, language=text, caption={The symbolic DA vector of $r^2$}, label={output:sda_r2}]
 Base  [ 10 / 20 ]        V [29]  
 -------------------------------------------
 0 0 0     0    x**2 + y**2 + z**2
 1 0 0     1    2*x
 0 1 0     2    2*y
 0 0 1     3    2*z
 2 0 0     4    1
 0 2 0     7    1
 0 0 2     9    1
\end{lstlisting}
\bigskip
\begin{lstlisting}[style=mystyle, language=text, caption={The simplified symbolic DA vector of $1/r$}, label={output:sda_r_inv}]
 Base  [ 10 / 20 ]        V [29]  
---------------------------------------------
0 0 0     0    1.0/r
1 0 0     1    -1.0*x/r**3
0 1 0     2    -1.0*y/r**3
0 0 1     3    -1.0*z/r**3
2 0 0     4    1.5*x**2/r**5 - 0.5*r**(-3)
1 1 0     5    3.0*x*y/r**5
1 0 1     6    3.0*x*z/r**5
0 2 0     7    1.5*y**2/r**5 - 0.5*r**(-3)
0 1 1     8    3.0*y*z/r**5
0 0 2     9    1.5*z**2/r**5 - 0.5*r**(-3)
3 0 0    10    -2.5*x**3/r**7 + 1.5*x/r**5
2 1 0    11    1.5*y/r**5 - 7.5*x**2*y/r**7
2 0 1    12    1.5*z/r**5 - 7.5*x**2*z/r**7
1 2 0    13    1.5*x/r**5 - 7.5*x*y**2/r**7
1 1 1    14    -15.0*x*y*z/r**7
1 0 2    15    1.5*x/r**5 - 7.5*x*z**2/r**7
0 3 0    16    -2.5*y**3/r**7 + 1.5*y/r**5
0 2 1    17    1.5*z/r**5 - 7.5*y**2*z/r**7
0 1 2    18    1.5*y/r**5 - 7.5*y*z**2/r**7
0 0 3    19    -2.5*z**3/r**7 + 1.5*z/r**5
\end{lstlisting}
\end{minipage}
\end{listing}
\vspace{1em}

We can simplify both the computation and its outcome by introducing a new variable 
$r$ because computing the DA/SDA vector does not require any actual differentiation. In standard symbolic computation, defining a new variable would force us to find its derivatives with respect to the original variables and then apply the chain rule to arrive at the final expression. Whether this reduces complexity depends on the form of the function being differentiated. In contrast, within SDA calculations, whenever a complicated expression appears repeatedly, we can always replace it with a new symbol to streamline the final result. 

Obtaining explicit expressions as polynomials for all the derivatives brings us another advantage. We can calculate any specific derivative without calculating the lower-order ones. While using the numerical DA to obtain a specific derivative of order $n$, one has to calculate the DA vector up to order $n$, which includes all the terms from order 0 to $n$. The proposed method avoids these redundant computations. Furthermore,  within the expression of a derivative,  the terms in the polynomial do not depend on each other. The independence of the derivatives and the terms makes it easy to parallelize the calculation, should it be necessary to further improve the efficiency. 

\begin{table}[b]
  \caption{Number of terms in the calculation of derivatives}
  \label{tab:sda_terms}
  \centering
  \begin{tabular}{cc@{\hspace{10mm}}cccc@{\hspace{10mm}}cccc}
  \toprule
  \multicolumn{1}{c}{} & &\multicolumn{4}{c}{\hspace{-7mm}$f(x,y,z)$} & \multicolumn{4}{c}{\hspace{-3mm}$f(x,y,z,r)$} \\
  order                & $N_\mathrm{d}$  & $N_\mathrm{t}$  & Max  & Min & Avg &  $N_\mathrm{t}$  & Max  & Min  & Avg  \\ \midrule
  1                    &  3   &  12   &  4    &   4  &   4  &  3    &   1   &   1   &  1        \\
  2                    &  6   &   72  &  14    &   10  &  12   &   9   &   2   &    1  &    2      \\
  3                    & 10   &   190  &  20    &  10   & 19    &  19    &   2   &   1   &    2      \\
  4                    & 15    &  525   &  49    &  29   & 35    &  39    &  4    &   2   &     3     \\
  5                    & 21    &  1149   &  67    &  38   & 55    &  69    &   4   &   2   &      3    \\
  6                    & 28    &  2339   &  149    &  63   & 116    &   119   &    8  &  3    &      4    \\
  7                    & 36    &  4185   & 176     & 69    & 116    &   119   &  8    &   3   &     4     \\
  8                    & 45    &  7980   & 300     & 118    &  177   &   189   &  8    &   3   &      5    \\
  9                    & 55    &  14492   &  390    & 148    & 263    &   294   &   12   &    4  &       7   \\
  10                   & 66    &   24876  &  666    & 203    &  376   & 630     & 18     & 5     & 10        \\ \bottomrule
  \end{tabular}
  \end{table}

\section{Numerical experiments}
We have run numerical experiments to benchmark the proposed method using SDA against cppTPSA, GTPSA \cite{deniau2015generalised}, and ForwardDiff \cite{revels2016forward}. We developed a Python parser, with the help of ChatGPT 4o, to analyze the SDA output and generate C++ code that calculate the Taylor expansion and the derivatives of the function $1/r$ up to the 10th order. A simple wrapper was created to make the C++ functions callable in the Julia programming environment. \cite{bezanson2012julia, bezanson2017julia, julia2024} Both cppTPSA and GTPSA perform numerical DA/TPSA calculations. The SDA code was developed based on cppTPSA and uses the exact same algorithms. GTPSA is a different library in C++, and it has also been ported to Julia.\cite{Signorelli2024GTPSAjl} We wrote C++ code to benchmark against cppTPSA and Julia code to benchmark against GTPSA in our experiments.  All the numerical results presented in this section are carried out on a PC with an Intel i7-4820K CPU running at 3.70 GHz and 48 GB RAM.

To validate the SDA code, we calculate the Taylor expansion up to the 10th order of $1/r$ for 10,000 groups of \textit{x}, \textit{y}, and \textit{z}. The values of them are generated as random numbers with uniform distribution ranging from 0 to 10. The coefficient of each term in the Taylor expansion is compared with the respective one calculated by cppTPSA or GTPSA and the relative error is calculated as $|c-c_b|/|c_b|$, where $c$ is the coefficient and $c_b$ is the value that was benchmarked against. For all the coefficients in each order of the Taylor expansion, we record the one with the maximum relative error in all the 10,000 Taylor expansions. The records are shown in Table~\ref{tab:validation}. Under the respective table header, cppTPSA or GTPSA, the left column shows the maximum relative error for each order listed, and the right column shows the value of the coefficient.  We can see the proposed method agrees well with both cppTPSA and GTPSA. While the relative error increases as the order goes up, the value is below $1.5\times 10^{-9}$ until order 10. The absolute error actually decreases since the absolute value of the coefficients tends to decrease drastically with the order.

\begin{table}[t]
  \caption{Validation of the SDA code}
  \label{tab:validation}
  \centering

  \begin{tabular}{c@{\hspace{2mm}}S[table-format=1.3e2] @{\hspace{7mm}} S[table-format=1.3e2]
    @{\hspace{8mm}}S[table-format=1.3e2] @{\hspace{7mm}} S[table-format=1.3e2]
    @{\hspace{8mm}}S[table-format=1.3e2] @{\hspace{7mm}} S[table-format=1.3e2]}
  \toprule
  \multicolumn{1}{c}{} & \multicolumn{2}{c}{cppTPSA} & \multicolumn{2}{c}{GTPSA} & \multicolumn{2}{c}{ForwardDiff} \\
  order                & \text{Error}  & \text{Value}  & \text{Error} & \text{Value} & \text{Error} &  \text{Value} \\ \midrule
  1                    & 4.441e-15   &  -2.263   &  7.169e-16    &   -2.420e-3  &   1.365e-16  &  -1.271e-3       \\
  2                    &  1.752e-10   &  6.683e-10&  2.550e-12    &   -6.377e-8  &  9.015e-16   &   -6.157e-3       \\
  3                    & 5.135e-12  &   1.531e-9  &  1.587e-10    &  1.793e-9  & 2.754e-16   &  -3.937e-4    \\
  4                    & 2.036e-11    &  7.851e-11  &  2.084e-11    & 3.902e-8    & 1.184e-16    &  -9.376e-1   \\
  5                    & 1.307e-10    &  -5.489e-12   &  4.510e-11    &  1.410e-8   & 8.037e-16    &  -3.868     \\
  6                    & 9.354e-11   &  -1.010e-12  &  5.457e-11    &  4.802e-10   & 1.986e-16    &   -1.092e-4    \\
  7                    & 2.144e-10    &  4.879e-13  & 2.110e-10     & 4.603e-11    & 3.676e-16    &   -7.732e1   \\
  8                    & 2.402e-10    &  1.482e-9   & 1.255e-9    & 5.152e-12    &     &     \\
  9                    & 1.482e-9    &  -9.737e-15  &  4.911e-9    & -3.056e-9    &     &     \\
  10                   & 1.431e-9    &  1.648e-13  &  1.087e-9   & -3.024e-8    &    &    \\ \bottomrule
  \end{tabular}
  \end{table}

  ForwardDiff is a popular Julia library for forward mode algorithmic differentiation. It has the best efficiency among similar algorithmic differentiation libraries for the first and the second order derivative calculation and is often used to benchmark new differentiation methods. Higher-order derivatives can be calculated recursively by taking the first order derivative against the lower-order derivative result. We calculated all the partial derivatives of $1/r$ up to the 7th order for 10,000 randomly generated \textit{x}, \textit{y}, and \textit{z}, and calculated the relative error with respect to the result by ForwardDiff. The ones with the maximum relative error in each order are presented in Table~\ref{tab:validation} under the table header ForwardDiff. The proposed method agrees with ForwardDiff very well. The relative errors are all in the order of $10^{-16}$. The absolute values of the higher-order derivatives are greater than the Taylor expansion coefficients due to the factor that is composed of the product of factorials.  
  
  High efficiency is always desired in large-scale scientific simulations. To check the performance of the proposed method, we compared its computational speed for $1/r$ with cppTPSA, GTPSA, and ForwardDiff. For the proposed method and cppTPSA, we calculated the average time for each order over the aforementioned 10,000 random samples. The calculations using GTPSA and ForwardDiff were carried out in Julia, so we used macros from the BenchmarkTools package \cite{JuliaCIBenchmarkTools} in Julia to measure their execution. For GTPSA, we ran the @belapsed macro \textemdash which returns the minimum elapsed time \textemdash 10 times and took the average. For ForwardDiff, we obtained the time from the output of the @btime macro. The computation time up to the 7th order for ForwardDiff and to the 10th order for all the others are listed in Table~\ref{tab:benchmark}. ForwardDiff shows good efficiency for the first- and the second-order derivatives, which is as expected. However, starting from the third order, ForwardDiff is significantly slower than the others, which is probably due to the recursive process in higher-order derivative calculation. cppTPSA and GTPSA show similar performance. cppTPSA is slightly faster up to the 5th order, while GTPSA overtakes from the 6th order onward. However, there is no order-of-magnitude difference between the two. Compared to the other three, the proposed method demonstrates a substantial advantage in speed. In all  orders, it takes less than 10\% of the computational time of any other library. For the other three libraries, all the lower-order terms have to be calculated before one can calculate a higher-order term. To have a fair comparison, we did the same thing in our code, but this is not necessary in practice. Since we have the explicit expression for each derivative, the calculations are totally independent of each other. For example, the computation time  is 2285 ns up to the 9th order and 3313 ns up to the 10th order. We can calculate only the 10th order derivatives and the time cost should be 1028 ns, the differentiation between the two previous numbers. Actually, we can calculate any specific derivative without calculating the others. The proposed method is dramatically faster than the others when up to a few derivatives are needed. 

\begin{table}[]
  \caption{Computation time for higher-order derivatives in nanoseconds}
  \label{tab:benchmark}
  \centering
  \begin{tabular}{c@{\hspace{10mm}}rrrr}
  \toprule
    order                & SDA  & cppTPSA  & GTPSA  & ForwardDiff \\ \midrule
  1                    &  42   &  793   &  1213    &   987      \\
  2                    &  70   &   920  &  1685    &   1630        \\
  3                    & 124   &   1462  &  2083    &  10200      \\
  4                    & 209    &  2079  &  2702    &  26800        \\
  5                    & 341    &  3720   &  3725    &  79700     \\
  6                    & 599    &  6657   &  5779   &  311100    \\
  7                    & 873    &  11860   & 9340     & 1847000   \\
  8                    & 1381    &  22817   & 15570     & -   \\
  9                    & 2285    &  36060   &  25350    & -     \\
  10                   & 3313    &   64164  &  41511    & -    \\ \bottomrule
  \end{tabular}
  \end{table}

\FloatBarrier

\section{Summary and future work} \label{sec:summary}
We presented a new method for automatic higher-order differentiation, which combines algorithmic differentiation with symbolic calculation using our SDA library. The new method allows us to obtain explicit expressions for all the derivatives up to a predetermined order using algorithmic computation. The expressions can then be used to generate code in a target programming language for differentiation. The expressions can be easily simplified by replacing repeatedly occurring complex terms with new variables, which helps to improve the efficiency of the differentiation code. Numerical experiments show that the differentiation code based on SDA results agrees with other libraries to machine precision, but only takes a fraction of time to run. It can also calculate a specific dderivative, avoiding redundant calculation of the lower-order ones.  

At present, the two steps of obtaining the derivative expressions using SDA, and generating code from those expressions, are carried out separately. In the future, we hope to fully automate these two steps, minimizing human intervention and making this method more user-friendly. The metaprogramming capabilities of the Julia programming language, which allow for dynamic code generation and execution, may help us achieve this goal.

\section*{Acknowledgement}
This material is based upon work supported by the U.S. Department of Energy, Office of Science, Office of Nuclear Physics under contract DE-AC05-06OR23177.

\printbibliography

\end{document}